\documentclass{cta-author}
\usepackage{times}
\usepackage{fancyhdr,graphicx,amsmath,amssymb}
{}
{}
{}
\usepackage[dvipsnames]{xcolor}
\usepackage{ragged2e}
\usepackage{lipsum}
\usepackage{algpseudocode}
\usepackage{algorithm}
\include{pythonlisting}
\usepackage{bibentry}
\usepackage{multirow}
\usepackage{array}
\usepackage{booktabs}
\usepackage{graphicx}
\usepackage{caption}
\usepackage{subcaption}
\usepackage[export]{adjustbox}
\usepackage{tabularx}
\usepackage{booktabs}
\usepackage{tikz}
\usepackage{lipsum}

\newcommand\copyrighttext{%
  \footnotesize This paper is a preprint of a paper accepted by IET Generation Transmission and Distribution
and is subject to Institution of Engineering and Technology Copyright. When the final version is published,
the copy of record will be available at the IET Digital Library}
\newcommand\copyrightnotice{%
\begin{tikzpicture}[remember picture,overlay]
\node[anchor=south,yshift=-5pt] at (current page.south) {\fbox{\parbox{\dimexpr\textwidth-\fboxsep-\fboxrule\relax}{\copyrighttext}}};
\end{tikzpicture}%
}
\begin{document}

\supertitle{Research Article}

%\title{Real-time Wide Area Voltage Stability Analysis with Consideration of Q-limits and N-1 Contingencies}

\title{Hybrid Voltage Stability and Security Assessment using Synchrophasors with Consideration of Generator Q-limits} 

\author{\au{Syed Muhammad Hur Rizvi$^{1\corr}$}, \au{Pratim Kundu$^{1}$}, \au{Anurag K. Srivastava$^{1}$}}

\address{\add{1}{Department of Electrical Engineering and Computer Science, Washington State University, Pullman, U.S.A.}
\email{syed.rizvi@wsu.edu}}

\begin{abstract}
Increased power demands, push for economics and limited investment in grid infrastructure have led utilities to operate power systems closer to their stability limits. Voltage instability may trigger cascade tripping, wide-area voltage collapse and power blackouts. Real-time voltage stability monitoring possible with deployment of phasor measurement units (PMUs) is essential to take proactive control actions and minimize the impact on system. This paper presents a novel online algorithm for a) hybrid perturbation analysis-based voltage stability monitoring (HPVSM), b) including Q-limit in voltage stability index and c) real time security analysis using voltage stability index. HPVSM based voltage stability index is computed using the data obtained from linear state estimator and PMU measurements. Typically measurement-based schemes ignore the impact of generator Q-limits and security analysis is not feasible. The proposed HPVSM based index considers the impact of generator Q-limits violations by anticipating the critical generators using real-time PMU measurements. Contingencies are ranked using the proposed voltage stability index for security analysis. Results simulated for the 9-bus WECC, IEEE 14, 57 and 118 bus systems highlights the superiority of the proposed method in real time voltage stability and security analysis.
\end{abstract}

\maketitle
\copyrightnotice
\section{Introduction}\label{sec1}
Power system voltage stability assessment is performed to identify impending instability condition \cite{Taylorstability} using knowledge of existing power system state. Traditionally voltage stability has been analyzed using continuation power flow (CPF) \cite{ajjarapu1992continuation,IETCont1,IETCont2}. CPF approaches because of their computational requirements are suitable for planning studies rather than real-time monitoring. Phasor Measurement Units (PMUs) provides opportunity for real-time stability monitoring schemes \cite{phadkePMUapp}. \par
Machine learning based approaches have also been used for synchrophasor assisted voltage stability assessment. Decision tree based approach has been presented in \cite{ML1decisiontrees}. In \cite{ML2regtree}, Zheng et. al. proposed regression tree based real-time voltage stability assessment where robustness against measurement errors and topology variation is also analyzed. Artificial neural network based approach has also been used for voltage stability assessment \cite{MLNN1,MLNN2,MLNN3}. A Major issue with machine learning based approach is lack of actual labelled training data leading to overly optimistic results and dynamic nature of the power system operation with multiple operating points.   \par
Local voltage and current phasors are used for obtaining the classic Thevenin equivalent for voltage stability assessment
\cite{VUthevenin,corsithevenin,LiIET1}. 
The ratio of the load and Thevenin impedance is used as an index for real-time voltage stability monitoring. As the system approaches instability, load impedance and Thevenin impedance become almost equal in magnitude.  Thevenin parameters are obtained by solving an over-determined system of equations using a window of real-time PMU measurements at the concerned bus. The Thevenin based voltage stability assessment concept has been extended to wide area voltage stability monitoring in \cite{Milosevicthevenin, Zimathevenin}. Coupled single port theory has also been used for wide area voltage stability monitoring \cite{coupled1,coupled2,coupled4,IET4}. One problem associated with measurement based techniques based on Thevenin equivalent is convergence issues because of very little change in operating conditions during normal operation \cite{comp1}.\par
Inability of the traditional measurement based approaches to consider the impact of generator reactive power limits has been a hindrance in efficient voltage stability monitoring. Recent work focused on possible extension of measurement based stability analysis techniques to consider the impact of reactive power limits \cite{matavalam2017sensitivity} using the concept of continuation power flow predictor, but this approach doesn't consider architecture for real-time analysis. A correction factor is introduced in \cite{DalaliIET3} to predict the impact of Q-limits on Thevenin-based voltage stability index, but this work doesn't consider the mathematical formulation to estimate the impact of Q-limits. Impact of Q-limits on security analysis is not considered in \cite{matavalam2017sensitivity,DalaliIET3}. Impact of Q-limits on coupled single port based approach is presented in \cite{su2015estimating}. A hybrid approach proposed in \cite{yuan2015hybrid} considers security analysis, but the impact of Q-limits on the index is not considered. A measurement based approach presented in \cite{HuIET2} considers the impact of n-1 contingencies on power limits of individual tie-lines without considering the impact of Q-limits. There is a need for a real-time voltage stability assessment method which can perform monitoring with consideration of reactive power limits and provides operators a feature to analyze what-if contingency scenarios. Moreover, impact of estimation issues on assessment performance (e.g. traditional Thevenin based approaches) should be minimized.  \par
%The proposed approach considers the impact of generator Q-limits in a hybrid manner which resolves the estimation issues. Moreover the proposed approach also facilitates online security monitoring.\parThe novelty of the proposed approach lies in the consideration of Q-limits in the VSI (Voltage Stability Index) and online security assessment
This paper introduces a hybrid perturbation analysis based voltage stability monitoring (HPVSM) approach. The proposed approach is considered hybrid because it utilizes real-time PMU measurements and system network model. network model is used by linear state estimator to provide best estimate of wide-area PMU measurements. Linear state estimator implementation considered by \cite{LSEBose0,LSEBose1} can enhance the quality of signals used for the proposed application. The novelty of the proposed approach lies in the consideration of Q-limits in the Voltage Stability Index (VSI) and online security assessment to suggest proactive control. The main contributions in this work are summarized as:
\begin{itemize}
    \item Consideration of reactive power limits in the computation of the voltage instability index using the proposed hybrid approach.
    \item A new index is proposed that combines the two causes of voltage instability i-e transmission network limit and reactive power limit in one index.
    \item Online security analysis for contingency screening and ranking, with consideration of reactive power limits is proposed. 
    \item Validation with multiple test cases. 
\end{itemize}
%Paper is organized as follows. Section II provides background theory and briefly explains proposed hybrid approach. Section III discusses consideration of reactive power limit and Section IV discusses online security analysis. Section V presents case studies and Section VI provides concluding remarks.
%%\vspace{-0.5cm}
\section{Background on Measurement based and Hybrid Voltage Stability Assessment}\label{sec2}
Measurement based voltage stability approaches have attracted lots of attention due to the increased deployment of Phasor Measurement Units in the modern power grid. The concept of assessing voltage stability using synchophasor data was first realized in the Thevenin based voltage stability index. Despite the apparent advantages of measurement based stability, the implementation issues became apparent mainly because of estimation problems. These issues highlight the need to move towards hybrid voltage stability assessment approaches where the assessment utilizes both measurement data and power system model information. Hybrid schemes have the potential to be more robust to estimation issues and are especially attractive for long-term but real time voltage stability assessment. %This section discusses the Thevenin based voltage stability approach and the concept of hybrid approach as a background for our proposed approach.
%%\vspace{-0.5cm}
\subsection{Thevenin-based Voltage Stability Index}
Use of local phasor measurements for construction of the Thevenin equivalent at the concerned bus is proposed in \cite{VUthevenin}. Least squares estimation is performed on a window of phasor data for determination of Thevenin parameters (Thevenin Impedance and Voltage). It is assumed that system conditions remain almost the same during the window. 
\begin{align*}
      Z_{th}=r+j.v, E_{th}=u+j.w
  \end{align*}
  Here, $Z_{th}$ is the Thevenin impedance, $r$ and $v$ are the real and imaginary parts of $Z_{th}$. $E_{th}$ represents the Thevenin voltage, and $u$ and $w$ are its real and imaginary parts. Using the complex voltage and current measurements at the load bus an over-determined set of equations can be set up \cite{VUthevenin} for finding the Thevenin parameters. The Voltage stability index is computed as follows:
      \begin{equation}
      VSI=\frac{|Z_{th}|}{|Z_{load}|}, Z_{load}=\frac{V_{(load)}}{I_{(load)}}
  \end{equation}
  %%%\vspace{-0.3cm}
 % \begin{equation}
  %    Z_{load}=\frac{V^{(1)}}{I^{(1)}}
  %\end{equation}
Despite its apparent simplicity and ease of interpretation this approach suffers with estimation issues. Accurate estimation of Thevenin parameters is not guaranteed if the change in measurement set is not enough. Moreover, the assumption that system conditions remain same during the estimation window can lead to inaccurate results.
%%\vspace{-0.4cm}
\subsection{Hybrid Voltage Stability Index Computation}
The Hybrid approach uses PMU measurements filtered by linear state estimator along with power system model information to compute VSI for all buses using just one set of measurements \cite{biswas2018voltage,biswas2}. \par

\subsubsection{Perturbation Analysis and Computation of VSI}
Perturbation for computation of VSI is modelled by changing the real and reactive power injections as per the load change direction. Applying perturbation on the Jacobian, fictitious measurements are obtained as follows:
 \begin{equation}
\begin{bmatrix}
\Delta \delta\\ 
\Delta V\\
\end{bmatrix}=\\
\begin{aligned}
\begin{bmatrix}
J_{p\theta}&J_{pv}\\ 
J_{q\theta}&J_{qv}\\
\end{bmatrix}^{-1}
\begin{bmatrix}
\Delta P\\ 
\Delta Q\\
\end{bmatrix}
\end{aligned}
\end{equation}
%%\vspace{-0.1cm}
\begin{equation}
\begin{bmatrix}
 \delta^f\\ 
|V|^f\\
\end{bmatrix}=\\
\begin{bmatrix}
\Delta \delta +\delta\\
\Delta V + |V|
\end{bmatrix}
\end{equation}

Using fictitious measurements Thevenin impedance is computed as follows:
  \begin{equation}
      Z_{th}=\frac{V^f-V}{I-I^f}
  \end{equation}
  Here measurement $V^f$ corresponds to the complex fictitious measurement obtained from perturbation analysis and $V$ corresponds to the complex voltage measurement obtained from linear state estimation results. Once the value of $Z_{th}$ has been computed, the VSI is obtained using (1).
  %%\vspace{-0.3cm}
  \section{Hybrid Perturbation Analysis based Voltage Stability Monitoring (HPVSM)}
In this paper a hybrid approach has been introduced that has the ability to accommodate Q-limits and perform online security analysis. The approach is considered hybrid because all the analysis is carried out based on the results of the linear state estimator. The proposed method predicts the critical generators and evaluates the impact of Q-limits on the voltage stability. The method computes a weighted voltage stability index, where weights are computed using PMU Data and the reactive power reserve information. Security analysis feature in the proposed hybrid approach is used for online contingency screening. Ability to perform online security analysis can enhance the awareness of the operator and improve the system resilience. %In case of missing data in PMU stream for linear estimator different approaches existing in literature for missing data reconstruction should be adopted \cite{PMUMissing1,PMUmissing2}.
%%\vspace{-0.5cm}
\subsection{Consideration of Q-limits}\label{sec3}
Voltage Stability problems often stem from insufficient availability of reactive power. Thevenin based voltage stability index inherently does not have the ability to anticipate Q-limits and leads to optimistic estimation of the margin from collapse. In this section, a new method for anticipation of Q-limits is considered which is able to predict the critical generators, and then accommodate the impact directly in the computed voltage stability index.
\subsubsection{Prediction of the Critical Generators}
In order to consider the impact of generator reactive power limits, the reactive power reserve of each generator in the system is modelled as a quadratic function of total reactive power load consumption similar to \cite{karki2009methods} where reactive power loss in the network is modelled. It is assumed that current reactive power generation and reactive power consumption at all the load buses is available from PMU measurements. Let, $RPR_i$ be the reactive power reserve of $i_{th}$ generator and $Q_T$ be the total reactive power consumption in the system. $Q_T$ is considered equal to sum of reactive power consumption at all the load buses. The $i^{th}$ generator, reactive power reserve is modelled as follows:
%%\vspace{-0.1cm}
\begin{equation}
    RPR_i=a_iQ_T^2+b_iQ_T+c_i
\end{equation}

In order to identify the parameters weighted least squares algorithm can be used on window of synchrophasor data. In a simplistic scenario where window has 4 measurements over-determined system of equations for least squares estimation would take the following form:
%%\vspace{-0.2cm}
\begin{equation}
    \begin{bmatrix}
    RPR_i(n)\\
    RPR_i(n+1)\\
    RPR_i(n+2)\\
    RPR_i(n+3)
    \end{bmatrix}
    =
    \begin{bmatrix}
    Q_T^2(n) &Q_T(n)&1\\
    Q_T^2(n+1) &Q_T(n+1)&1\\
    Q_T^2(n+2) &Q_T(n+2)&1\\
    Q_T^2(n+3) &Q_T(n+3)&1
    \end{bmatrix}
    \begin{bmatrix}
        a_i\\
    b_i\\
    c_i
    \end{bmatrix}
\end{equation}
Parameters corresponding to all generators can be evaluated in the same way. \par
Once the parameters have been estimated the critical value of total load at which $RPR_i$ would become zero can be computed. This can be done by evaluating the roots of $RPR_i$ equation. The realistic roots can be selected to make a list of critical generators expected to hit the Q-limits. \par
%%\vspace{-0.5cm}
\begin{align}
    0=a_iQ_{cr}^2+b_iQ_{cr}+c_i\\
    Q_{cr}=\frac{-b_i\pm\sqrt{b_i^2-4a_ic_i}}{2a_i}
\end{align}

Once values of $Q_{cr}$ corresponding to each reactive power resource has been computed, list of most critical generators is identified, including ones having values close to $Q_{T}$. Threshold defined in Algorithm 1 is used to populate the list of critical generators. During thorough analysis it was observed that rather than having a fixed threshold for populating the list of critical generators, the threshold should be dependent on total system reactive power load. This approach makes the computed index less conservative. Time taken to compute $Q_{cr}$ for all generators of IEEE 57 Bus system is found to be 0.009206 seconds on Intel Core i-7 laptop computer when 30 measurements are used to set up over-determined system. 
\subsubsection{Evaluating the Impact of Q-limits}
In the proposed hybrid approach, we propose perturbation analysis by modification of the power system Jacobian. Power system Jacobian at the current operating point is given as: \par
%%%\vspace{-0.5cm}
  \begin{equation}
J=\\
\begin{aligned}
\begin{bmatrix}
J_{p\theta}&J_{pv}\\ 
J_{q\theta}&J_{qv}\\
\end{bmatrix}
\end{aligned}
\end{equation}
Let's assume that the generator at bus $x$ is predicted to hit the Q-limit next. Once it has been identified the Jacobian can be modified considering Bus $x$ as a PQ bus instead of PV bus.\par
%%%\vspace{-0.5cm}
  \begin{equation}
J'=\\
\begin{aligned}
\begin{bmatrix}
J_{p\theta}&J_{pv_x}&J_{pv}\\ 
J_{q_x\theta}&J_{q_xv_x}&J_{q_xv}\\
J_{q\theta}&J_{qv_x}&J_{qv}\\
\end{bmatrix}
\end{aligned}
\end{equation}

Here $v_x$ is the newly added variable which is voltage of generator $x$ and $q_x$ is the reactive power injection at Bus $x$. The size of the power flow Jacobian is increased by 1.

\subsubsection{Weighted Voltage Stability Index}
A weighted voltage stability index is introduced here. The main purpose is to combine the information received from voltage stability index computed at the current operating point and the anticipated voltage stability index with violation of Q-limits. 
 \begin{equation}
\begin{bmatrix}
\Delta \delta\\ 
\Delta V_x\\
\Delta V\\
\end{bmatrix}=\\
\begin{aligned}
\begin{bmatrix}
J_{p\theta}&J_{pv_x}&J_{pv}\\ 
J_{q_x\theta}&J_{q_xv_x}&J_{q_xv}\\
J_{q\theta}&J_{qv_x}&J_{qv}\\
\end{bmatrix}^{-1}
\begin{bmatrix}
\Delta P\\ 
\Delta Q_x\\
\Delta Q\\
\end{bmatrix}
\end{aligned}
\end{equation}

\begin{equation}
\begin{bmatrix}
 \delta^f\\ 
V_x^f\\
V^f\\
\end{bmatrix}=\\
\begin{bmatrix}
\Delta \delta +\delta^{SE}\\
\Delta V_x + V_x^{SE}\\
\Delta V + V^{SE}
\end{bmatrix}
\end{equation}

  \begin{equation}
      Z_{th}=\frac{V^f-V^{SE}}{I^{SE}-I^f}
  \end{equation}

    \begin{equation}
      VSI_{u}[i]=\frac{|Z_{th}[i]|}{|Z_{load}[i]|}
  \end{equation}
  
\begin{equation}
    WVSI[i]=w_1VSI[i]+w_2VSI_u[i]
\end{equation}
\begin{equation}
    w_1=\frac{\sum_{i\in List}RPR_i}{\sum_{i\in List} {Q_{max_i}}}
\end{equation}
\begin{equation}
    w_2=1-w_1
\end{equation}
Measurements with subscript $SE$ represent measurements filtered through Linear State Estimator. The weighted voltage stability index allows alarm processing by taking into account both current network load-ability constraints and status of reactive power reserves. Here the $List$ refers to a list of the identified critical generators using Algorithm 1. VSI represents the index computed without considering the impact of generator Q-limits, and $w_1$ is the corresponding weight. $VSI_u$ is the index computed assuming critical generator buses as PQ buses and $w_2$ is the corresponding weight.\par

%%\vspace{-0.0cm}
%\renewcommand{\labelenumi}{\Roman{enumi}}
\subsubsection{Algorithm}
Evaluation of the impact of Q-limits on the computed voltage stability index has two steps. The first step is prediction of critical generators expected to hit the Q-limits and the second step is estimating the impact on the computed voltage stability index at the current operating point.
%%\vspace{-0.3cm}
\begin{algorithm} [H]
%\SetAlgoLined
\begin{algorithmic}[1]
%\\\hrulefill
%\begin{enumerate}%{\roman{}}
\State Estimate the model parameters for Reactive power reserve of each generator as a quadratic function of load. 
%\makebox[\linewidth]{\rule{\columnwidth}{0.9pt}}
\State Estimate the roots of quadratic equation (05) to $Q_{cr}$ for the respective generator. Realistic roots are selected such that corresponding $Q_{cr}>$ $Q_T$ if the upper limit violation is concerned or $Q_{cr}<Q_T$ if the lower limit is concerned. Moreover, roots having imaginary parts are eliminated from the current list. %'\makebox[\linewidth]{\rule{\columnwidth}{0.9pt}}
\State Formulate a list of critical generators.  The criterion adopted in this study is to first select generator having minimum $Q_{cr}$ and then select the generators having $Q_{cr}$ in the range             $[Q_{cr}, Q_{cr}+th\times Q_T]$ to form the List. The threshold adopted in this study is 0.01. 
\State Convert the short listed generators to PQ mode and remodel the Jacobian matrix at the current operating point.
\State Use the modified Jacobian to evaluate the impact of nearest Q-limits in the form of weighted voltage stability index (WVSI).
%\end{enumerate}
\end{algorithmic}
\caption{Evaluating the Impact of Q-limits}
\end{algorithm}
%%\vspace{-0.5cm}
The concept is demonstrated for WECC 9 Bus system. The voltage stability index for the current operating point and anticipated index with consideration of reactive power limits is shown for Bus 7 as load is incremented until non-convergence of power flow. It can be seen in Fig. 1 that VSI jumps suddenly from 0.7 to more than 0.9 when system becomes unstable. This sudden jump comes due to reactive power reserve exhaustion of the second generator. Such a sudden jump can catch operator off-guard if relying on the VSI. WVSI, on the other hand anticipates the critical generators and estimates the impact of reactive power reserve exhaustion in advance, thereby providing operator an early alert and more time to devise preventive control actions.
\begin{figure}[htb]
% \begin{flushleft}
\centering
\includegraphics[width=\columnwidth, height=6.0cm ]{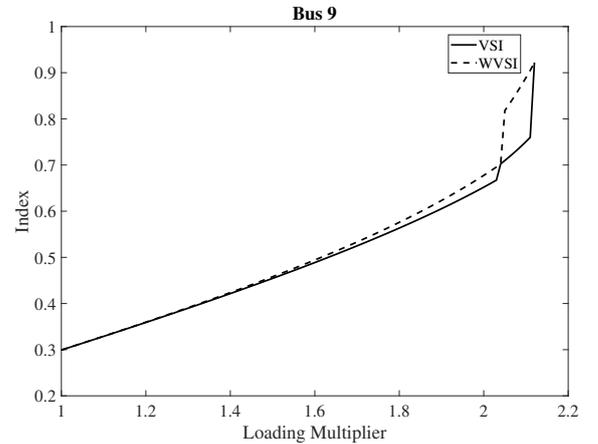}
\caption{\label{fig:ql}Anticipating Impact of Q-limits [WECC 9 Bus System]}
 %\end{flushleft}
\end{figure}
%%\vspace{-0.5cm}
\subsection{Security Analysis Using the Proposed Approach}\label{sec4}
This section introduces a hybrid approach to online security analysis. Traditional measurement based voltage stability approaches cannot analyze the impact of contingencies on the computed index in an online manner. Security analysis involves two steps for a given contingency, the first step is prediction of post contingency state and critical generators using the PMU measurements and network topology information. The second step involves computation of WVSI. The analyzed contingency is then classified as critical or non-critical. For the consideration of N-1 Contingencies piecewise linear method proposed in \cite{Sauercont} is adopted for prediction of post-contingent states and critical generators. 
\subsubsection{Piecewise Linear Estimation }
Piecewise linear and linear estimation of post-contingency states were considered in \cite{Sauercont}. Piecewise linear estimation is superior because of its ability to inherently consider the impact of Q-limits, which resulted in better estimation of voltage magnitudes and reactive powers of the generators. This subsection briefly reviews the concept behind the piecewise linear estimation.\par
Contingency to be analyzed is parametrized in terms of $K$ which varies from 0 to 1. Starting from iteration 1 (i=1) and piecewise linear parameter $K=0$, perturbation analysis is performed at each step to predict post-contingency states as $K$ varies from 0 to 1. Reactive power change at each unconstrained generator bus in terms of minimum ($   \underline{ \Delta Q_j^{(i)}}$) and maximum ($   \overline{ \Delta Q_j^{(i)}}$) reactive power generation allowed, at $i^{th}$ step is given as
\begin{align*}
       \overline{ \Delta Q_j^{(i)}}=\overline{Q_j^g}-Q_j^{g(i)}\geq 0\\
   \underline{ \Delta Q_j^{(i)}}=\underline{Q_j^g}-Q_j^{g(i)} \leq 0
\end{align*}
Voltage deviation ($\Delta V_j^{(i)}$) at each constrained bus for the $i^{th}$ is given as
\begin{equation*}
    \Delta V_j^{(i)}=V_j^{sp}-V_j^{(i)}
\end{equation*}
Critical constrained and unconstrained generator buses are identified. $P_Q$ and $N_Q$ represent critical unconstrained buses in terms of minimum and maximum reactive powers respectively. These lists are updated as $K$ varies from 0 to 1. $P_v$ and $N_v$ represent critical constrained generator buses in terms of minimum and maximum reactive powers respectively. Once the list is finalized step change in piecewise parameter $\Delta K$ is identified as
\begin{equation}
    \Delta K^{(i)}=\min {\{\min_{j \in P_Q} \frac{\overline{ \Delta Q_j^{(i)}}}{\frac{dQ_j^{(i)}}{dK}},\min_{j \in N_Q} \frac{\underline{ \Delta Q_j^{(i)}}}{\frac{dQ_j^{(i)}}{dK}}, \min_{j \in N_v U P_v}\frac{\Delta V_j^{(i)}}{\frac{dV_j^{(i)}}{dK}}\}}
    %\min_{j\epsilon P_Q}{\frac{\overline{ \Delta Q_j^{(i)}}}{\frac{dQ_j^{(i)}}{dK}},\min_{j\epsilon N_Q}{\frac{\underline{ \Delta Q_j^{(i)}}}{\frac{dQ_j^{(i)}}{dK}},\min_{j\epsilon N_v U P_v}{\frac{   { \Delta V_j^{(i)}}}{\frac{dV_j^{(i)}}{dK}}}
\end{equation}

Gradients,${\frac{dQ_j^{(i)}}{dK}}$ and ${\frac{dV_j^{(i)}}{dK}}$ are computed using mathematical formulation in \cite{Sauercont}. Once the value $\Delta K^{(i)}$ has been computed new value of $Q_j^{g(i)}$ and $V_j^{(i)}$ are computed using linear extrapolation for $\Delta K^{(i)}$. This is continued until $K$ approaches 1, which is the point which gives post contingency states.\par

\subsubsection{Contingency Screening}
In this work a list of contingencies is screened and critical contingencies are singled out. The first step is prediction of post contingency states. The second step is computation of post contingency voltage stability indices. The maximum voltage stability index of the system is used as screening index. The list of critical generators ($P_Q$ and $N_Q$) corresponding to the final state of piecewise linear estimation is used for computation of $WVSI$ for a selected contingency. From the list of contingencies contingencies having maximum WVSI are considered as most important contingencies. If any of the screened contingencies have threshold greater than a specified value like 0.75 then that contingency is considered as critical contingency. Most Thevenin based approaches use threshold in the range 0.7-0.9 for alarm processing. \par%Flow chart in Figure 3, summarizes the proposed method for contingency screening.\par
Fig. 2 shows the architecture of the proposed approach. Table I compares the proposed approach with Thevenin and previously proposed hybrid approach for Voltage stability index computation. Fig. 3 describes the architecture for the proposed security analysis approach.
%%\vspace{-.2cm}
  \begin{table}[h]
  \centering
  \caption{Comparison of Voltage Stability Monitoring Schemes}
\begin{tabular}{ccccc}
\toprule
Feature&Thevenin \cite{Milosevicthevenin,LiIET1,VUthevenin,corsithevenin} &Hybrid \cite{biswas2018voltage,biswas2} &Proposed\\
\cmidrule{1-4}
Range&0-1&0-1&0-1\\
\cmidrule{1-4}
Q-limits& & & \\
consideration&No&No&Yes\\
\cmidrule{1-4}
Threshold&$\approx 1$&  $\approx 1$&$\approx 1$\\ 
\cmidrule{1-4}
Security Analysis&No&No&Yes\\
\cmidrule{1-4}
Index Update &PMU&LSE&LSE\\
Rate& & & \\
\cmidrule{1-4}
Q-limits   &Not&Not&LSE\\
anticipation&Avail. &Avail. & \\
Update Rate& & & \\
\bottomrule
\end{tabular}
    \label{table:comp}
\end{table}

\begin{figure}[h]
 \begin{center}
\includegraphics[width=\columnwidth, height=8.0cm ]{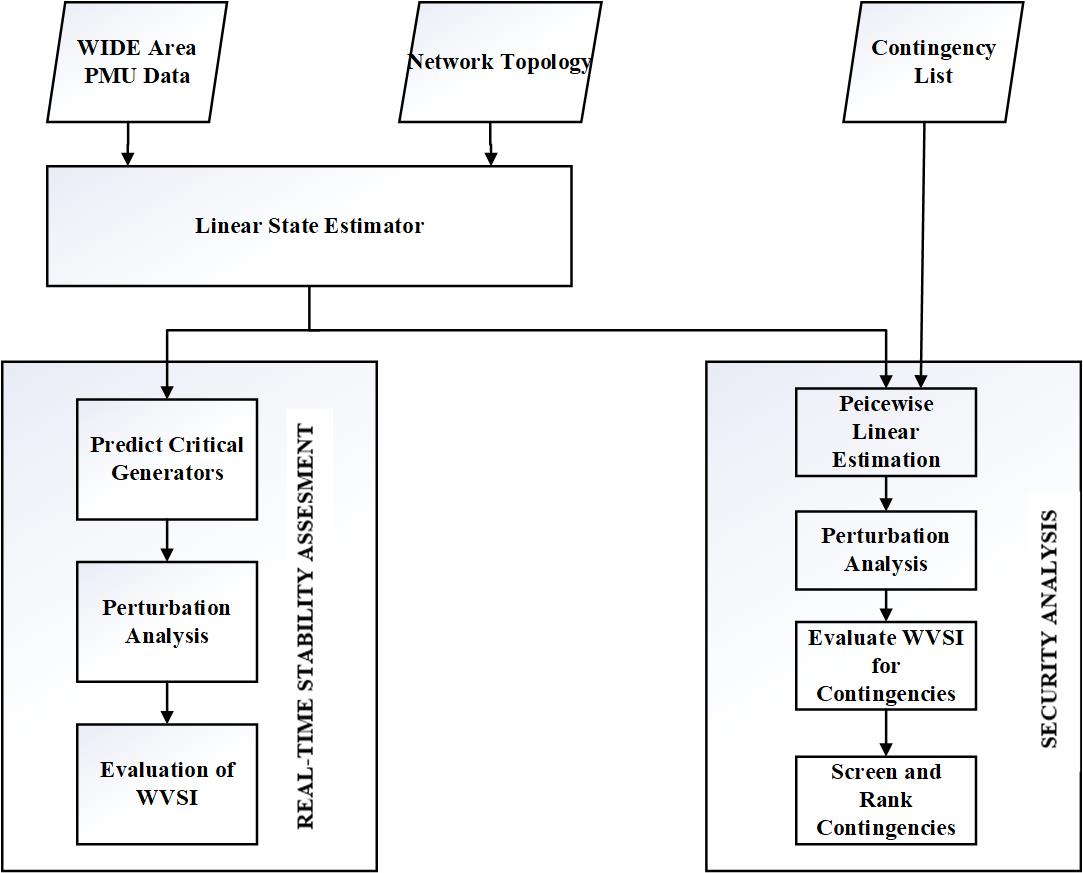}
\caption{\label{fig:hvsa}Hybrid Perturbation Based Voltage Stability Monitoring}
 \end{center}
\end{figure}

The advantage associated with this approach is that it is not prone to ill-conditioning issues associated with real-time least square based approaches. Moreover, proposed approach also avoids the assumption that system condition remains same during the estimation process. The update rate of the index using the proposed approach depends on the frequency of linear state estimator results. Since voltage stability is a slowly varying dynamic phenomenon, such a hybrid approach is fast enough for timely long-term voltage stability assessment. 
\begin{figure}[h]
 \begin{center}
\includegraphics[width=7cm, height=8cm ]{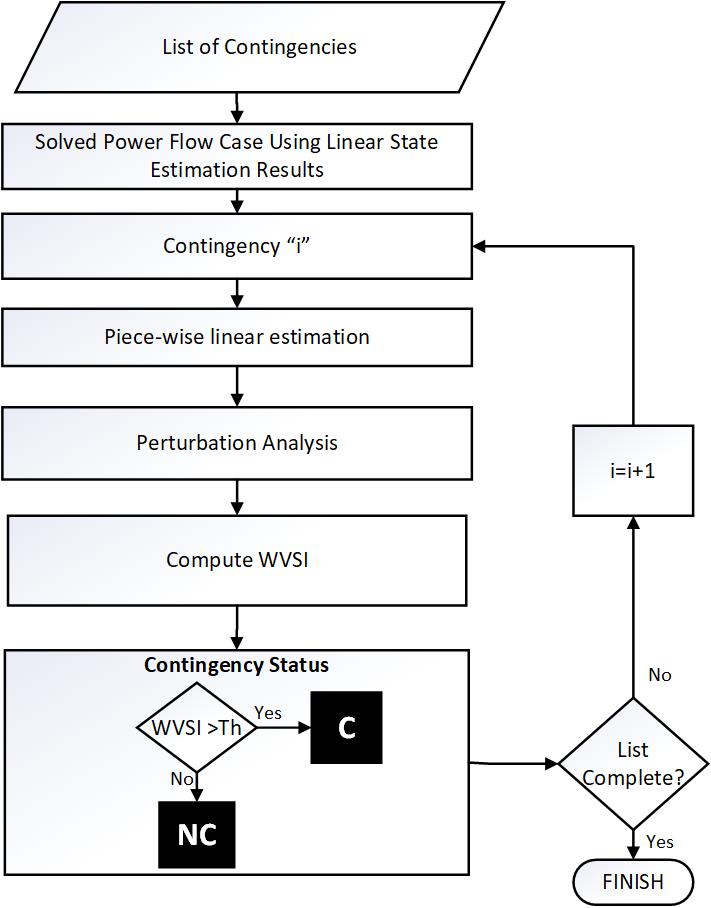}
\caption{\label{fig:ql}Security Analysis}
 \end{center}
\end{figure}
%%\vspace{-0.4cm}
\section{Case Studies and Simulation Results}\label{sec5}
This section presents case studies for voltage security analysis using the proposed method. IEEE 14 Bus, 57 Bus and 118 Bus systems have been studied. Subsection 4.1 provides the basis for the application of the considered hybrid voltage stability analysis method by validation from continuation power flow. In subsection 4.2 all the benchmark systems are studied with regards to anticipation of Q-limits. N-1 contingency screening results are presented in subsection 4.3.
%%\vspace{-0.4cm}
\subsection{Continuation Power flow based Validation}\label{subsec5.1}

In this subsection the effectiveness of the hybrid index computation method is validated using continuation power flow for different IEEE benchmark systems. Repeated power flow results are used to compute the index at each point as the load is varied from base case till power flow fails to converge. Continuation power results are then used for validation by comparison of PV curve and the WVSI for the most critical Bus. Continuation powerflow implementation in MATPOWER \cite{matpower} is used for validation.
\subsubsection{IEEE 14 Bus System}
For IEEE 14 Bus system load was increased from base case till power flow failed to converge. Corresponding to each converged solution voltage stability index was computed using the proposed method. Bus 9 was found to be the most critical bus from the observed values of Voltage stability indices just before collapse.\par
Continuation power flow was used to validate the results. Figure \ref{fig:14b} shows the PV curve of Bus 9 and the index value for same Bus. The proposed method was able to identify the voltage instability. Bus 14 and Bus 12 were also found to be critical with a threshold of 0.9 for IEEE 14 Bus system.  
\begin{figure}[t]
 \begin{center}
\includegraphics[width=\columnwidth, height=6.0cm ]{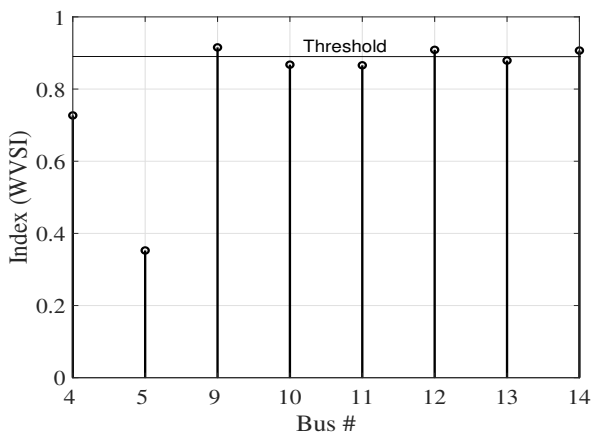}
\caption{  \label{fig:14a}VSI for Load Buses Just before Collapse [IEEE 14 Bus System]}
 \end{center}
\end{figure}

\begin{figure}[htp]
\centering
{\includegraphics[width=\columnwidth, height=6cm ]{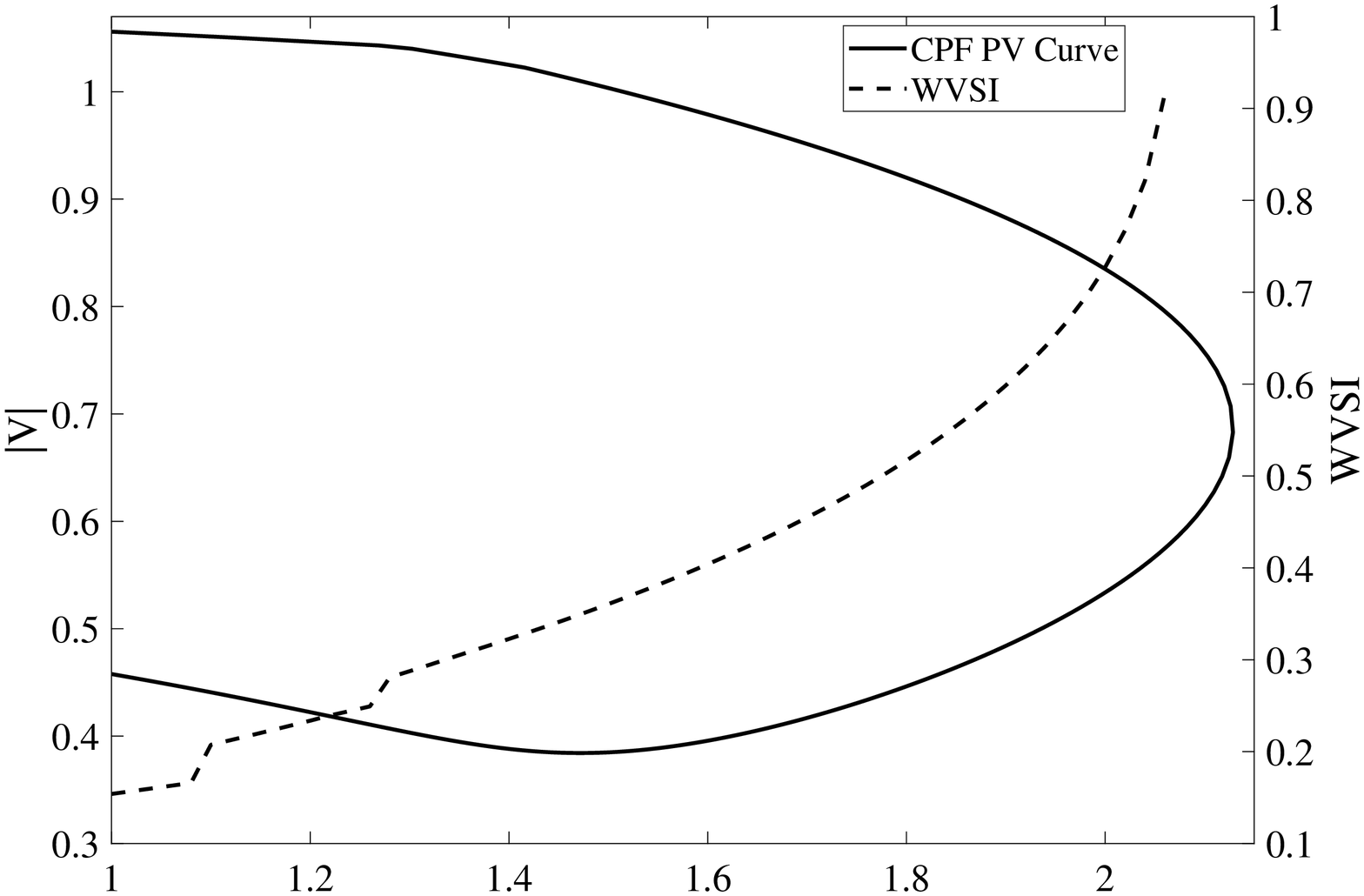}
\subcaption{\label{fig:14b}}}% Bus 9-IEEE 14 Bus System}}
{\includegraphics[width=\columnwidth, height=6cm ]{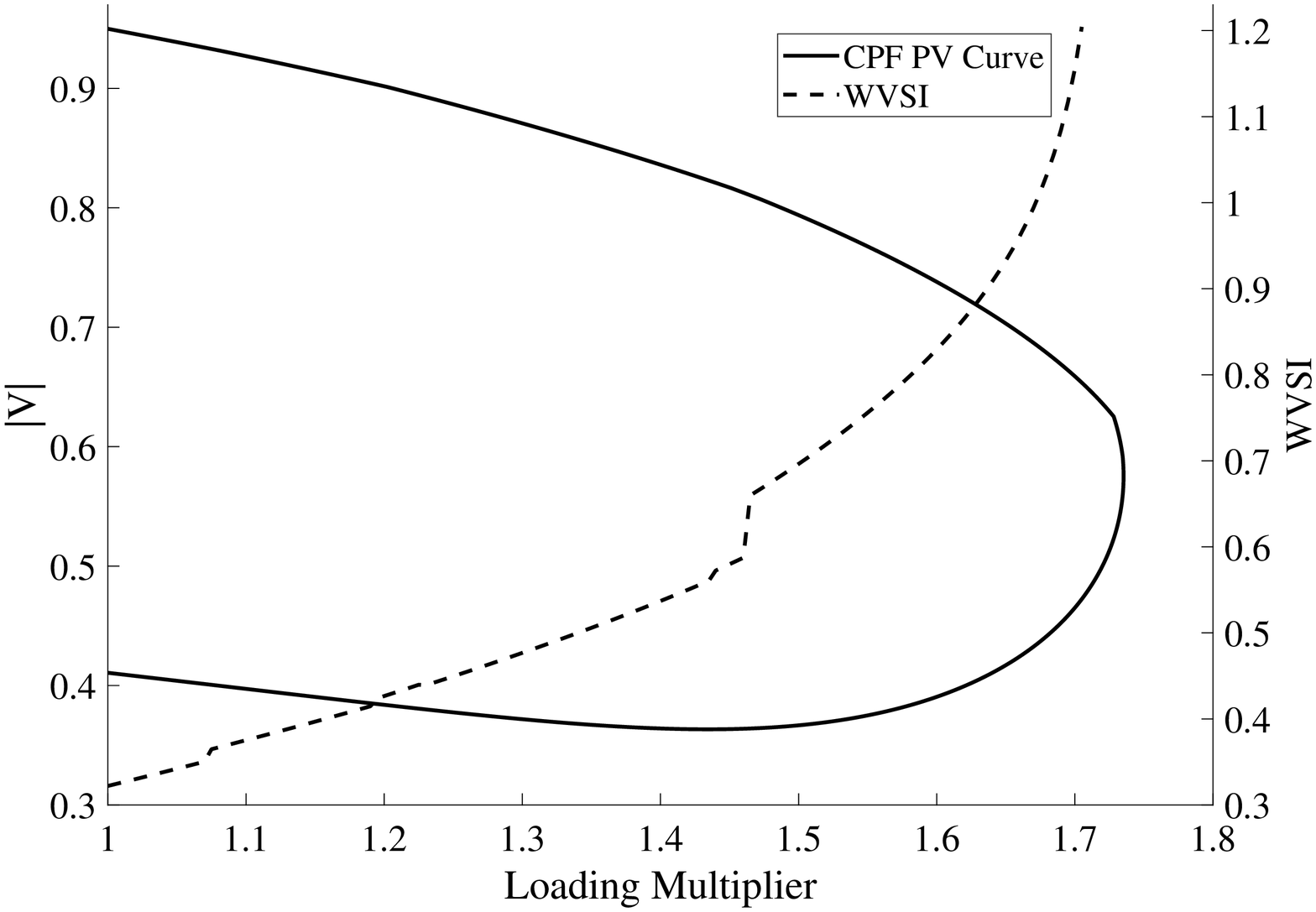}
\subcaption{  \label{fig:57b}}}% Bus 32-IEEE 57 Bus System}}
{\includegraphics[width=\columnwidth, height=6.0cm ]{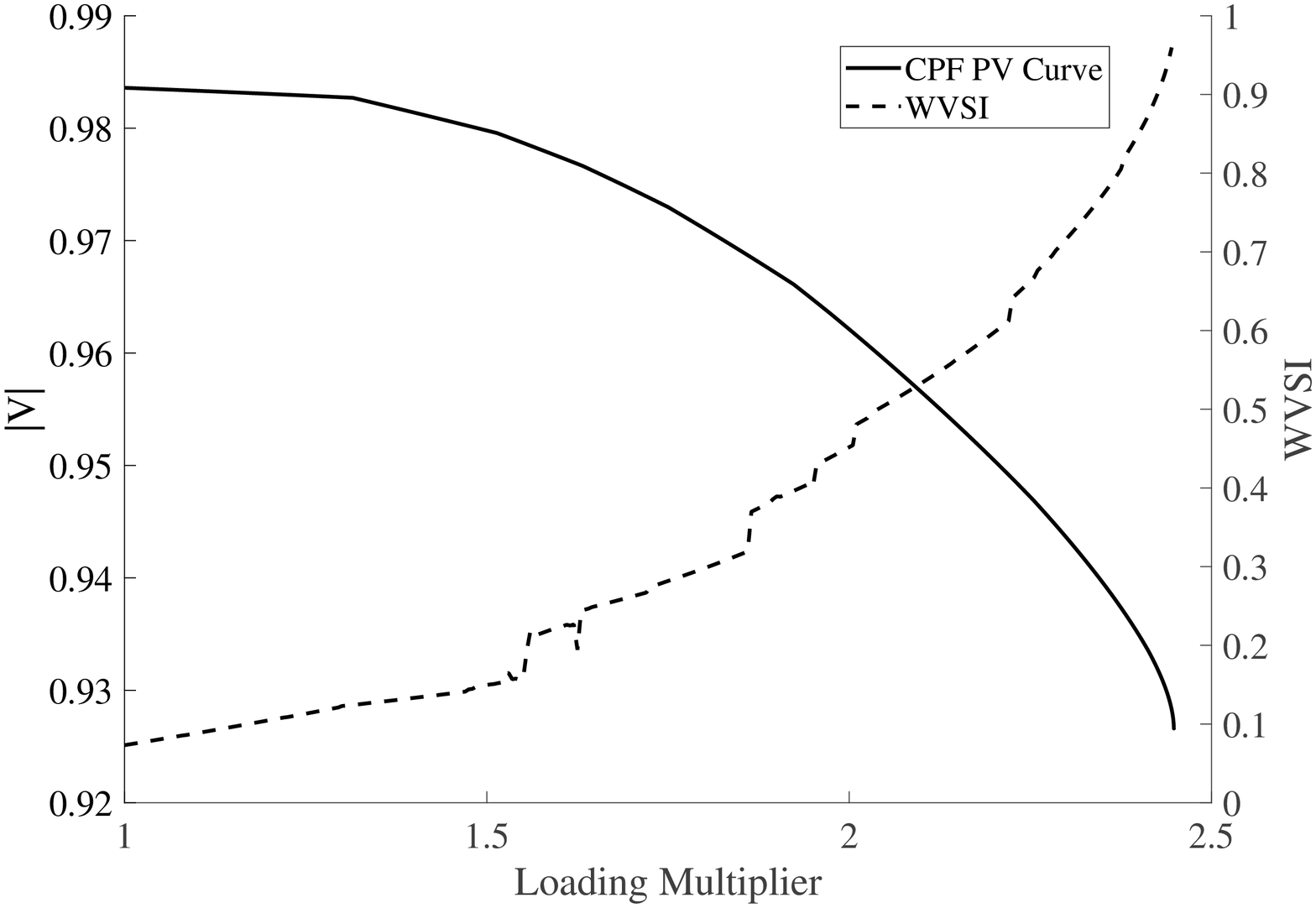}
\subcaption{\label{fig:118}}}% Bus 14-IEEE 118 Bus System}}
 \caption{CPF Based Validation for IEEE Benchmark Systems\figfooter{(a)}{IEEE 14 Bus System- Bus 9}
\figfooter{(b)}{IEEE 57 Bus System- Bus 32}
\figfooter{(c)}{IEEE 118 Bus System- Bus 14}}
\end{figure}
\subsubsection{IEEE 57 Bus System}
Bus 32 was observed to be the most critical bus from the observed values of Voltage stability indices just before collapse.Continuation power flow was then used to validate the results.Figure \ref{fig:57b} shows the PV curve of Bus 32 and the index value for same Bus. 
%\begin{figure}[t]
 %\begin{center}
%\includegraphics[width=9.cm, height=6.cm ]{IEEE57VSIbars2.eps}
%\caption{  \label{fig:57a}VSI for Load Buses Just before Collapse [IEEE 57 Bus %System]}
 %\end{center}
%\end{figure}

%\begin{figure}[h]
 %\begin{center}
 %%\vspace{-0.2cm}
%\includegraphics[width=9cm, height=6cm ]{CPFValid57.eps}
%\caption{  \label{fig:57b}CPF Based Validation [Bus 32-IEEE 57 Bus System]}
 %\end{center}
%\end{figure}
\subsubsection{IEEE 118 Bus System}
Similarly, for IEEE 118 Bus system load was increased from base case till power flow failed to converge. Corresponding to each converged solution voltage stability index is computed using the proposed method. Bus 14 was observed to be the most critical bus from the observed values of Voltage stability indices just before collapse.\par
Continuation power flow was used to validate the results. Fig. \ref{fig:118} shows the PV curve of Bus 14 and the index value for same Bus. The proposed method was able to identify the voltage instability. Buses 7, 14, 20, 29, 47 and Bus 60 were also critical if a threshold of 0.9 were selected for IEEE 118 Bus system.  
%\begin{figure}[h]
% \begin{center}
%\flushleft
%\includegraphics[width=9cm, height=6.0cm ]{cpfval118.eps}
%\caption{\label{fig:118}CPF Based Validation [Bus 14-IEEE 118 Bus System]}
 %\end{center}
%\end{figure}
%%\vspace{-0.5cm}
\subsection{Consideration of Reactive Power limits}\label{subsec5.2}
This subsection demonstrates the impact of reactive power limit consideration. First the generator with the $Q_{cr}$ closest to current total reactive load ($Q_T$) is selected in the list. Then list is updated to include generators having $Q_{cr}$ within the range [$Q_{cr}^{min}$, $ 0.01Q_T$+$Q_{cr}^{min}$]. Here $Q_{cr}$ is the predicted critical value of total load corresponding to the most critical generator. Once the list, comprising of the critical generators is populated, Algorithm 1 is used to accommodate reactive power limits in the index.

\subsubsection{IEEE 14 Bus System}
\begin{figure}[b]
\centering
 %\begin{center}
{\includegraphics[width=\columnwidth, height=6.0cm ]{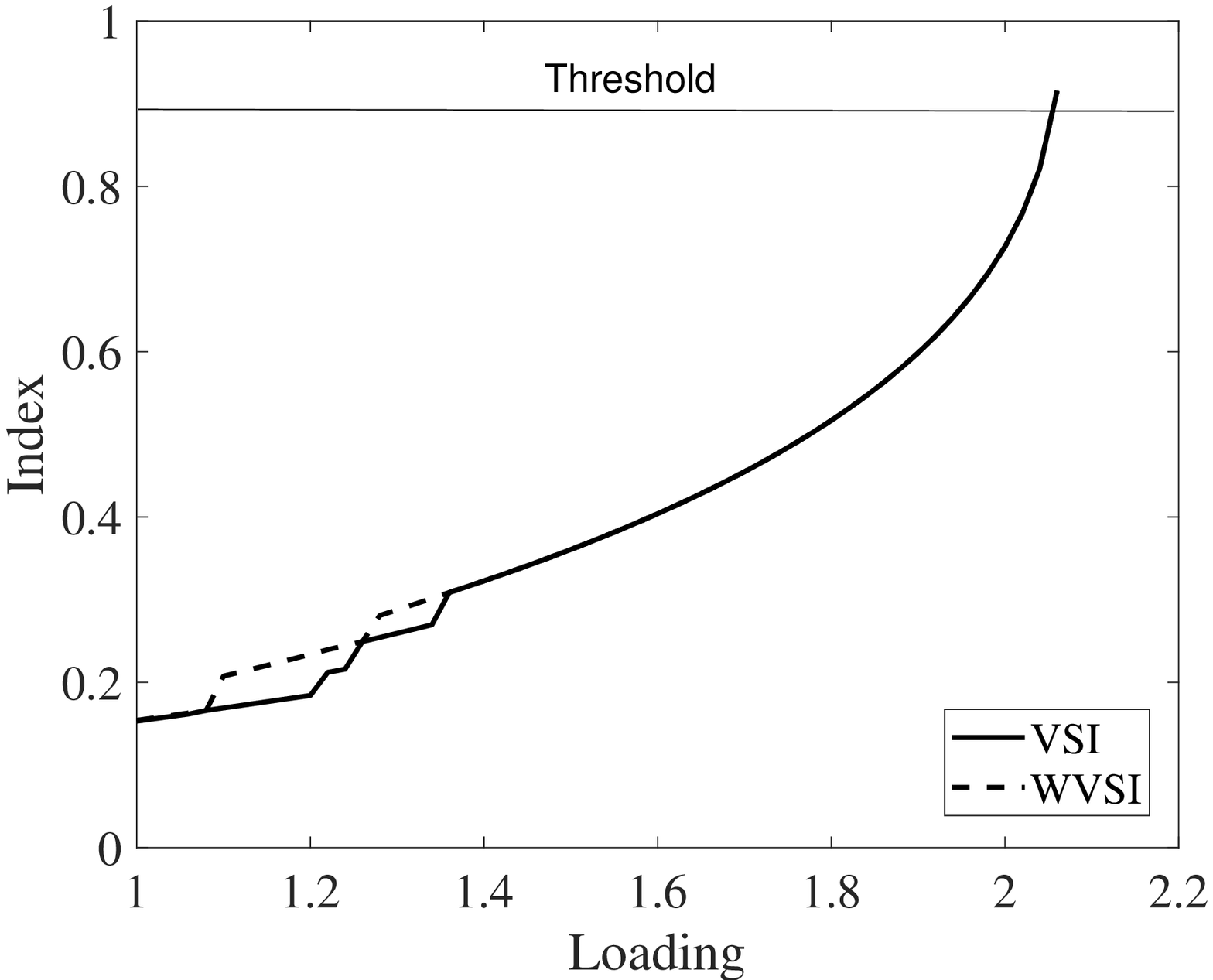}
\subcaption{  \label{fig:14wvsi}}}%Anticipating Impact of Q-limits [IEEE 14 Bus System]}
{\includegraphics[width=\columnwidth, height=6.0cm ]{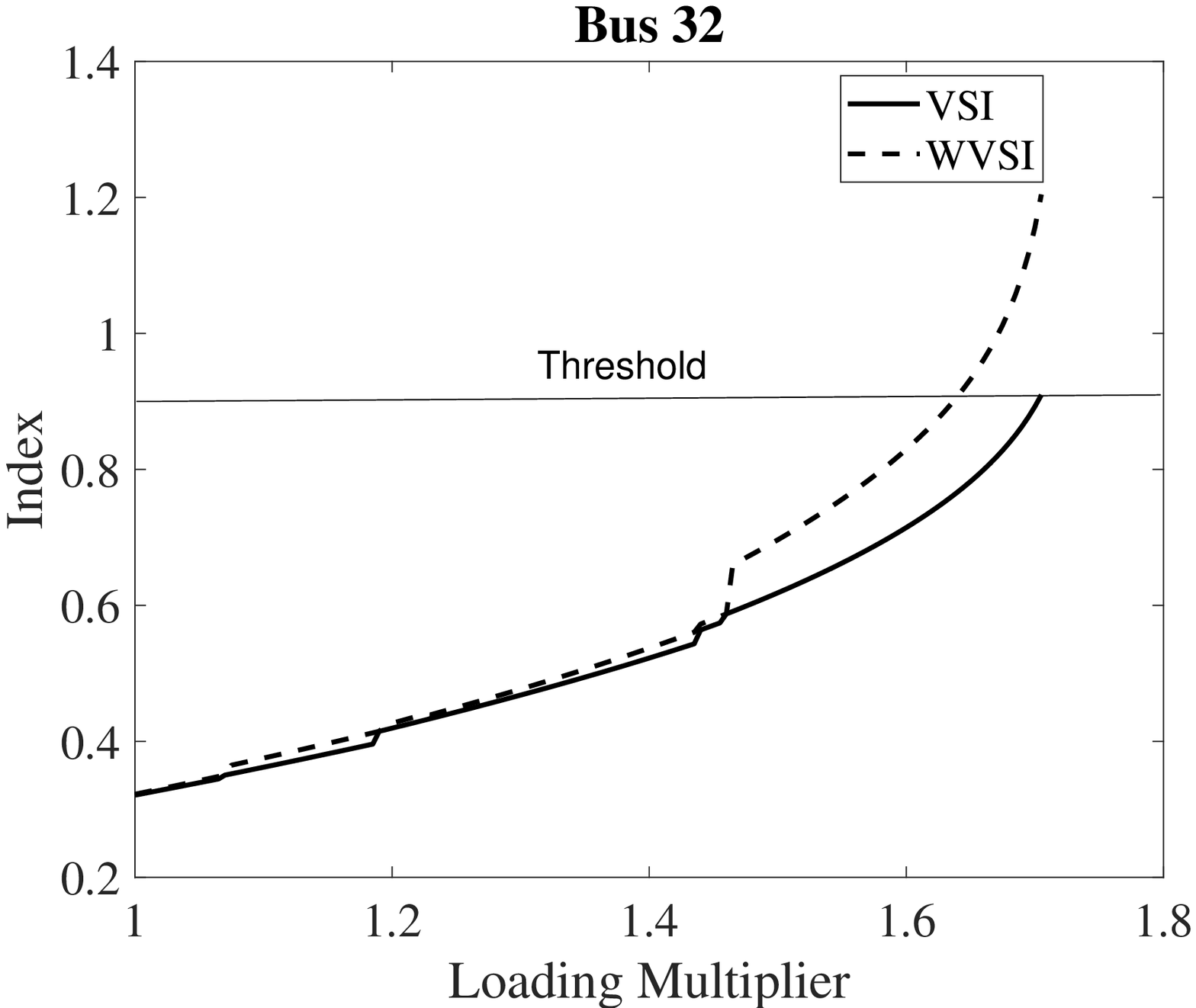}
\subcaption{  \label{fig:57wvsi}}}%Anticipating Impact of Q-limits [IEEE 57 Bus System]}
{\includegraphics[width=\columnwidth, height=6.0cm ]{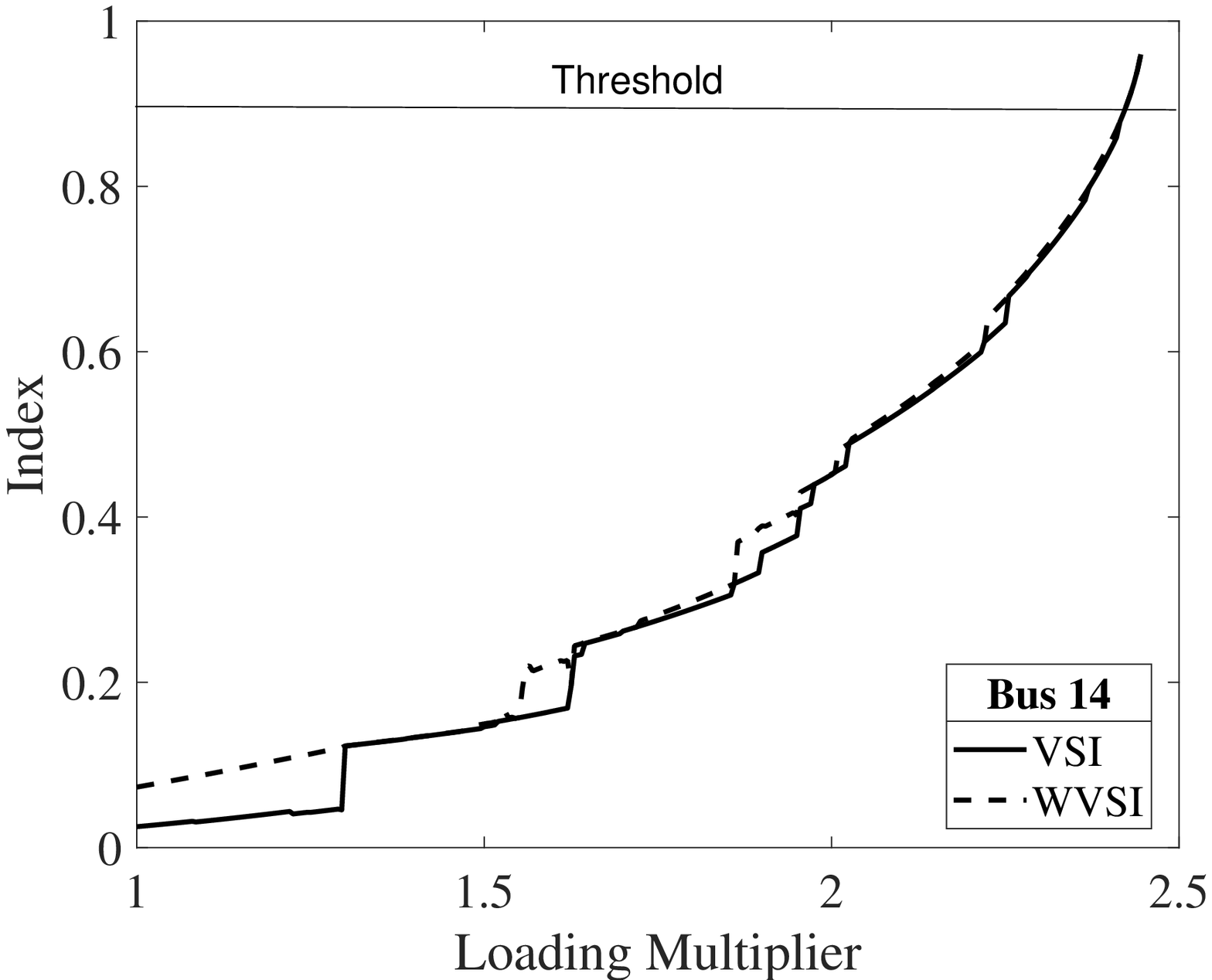}
\subcaption{  \label{fig:wvsi118}}}%Anticipating Impact of Q-limits [IEEE 118 Bus System]}
 %\end{center}
 \caption{Anticipating Impact of Q-limits for IEEE Benchmark Systems 
 \figfooter{(a)}{IEEE 14 Bus System- Bus 9}
\figfooter{(b)}{IEEE 57 Bus System- Bus 32}
\figfooter{(c)}{IEEE 118 Bus System- Bus 14}}
\end{figure}

IEEE 14 Bus system was studied using the proposed algorithm for identifying the critical generators and predicting the impact on the VSI. Near the base case loading generator 2 was the most critical generator closely followed by 8. At $k=$1.3, the generator at bus 2 was already exhausted in terms of reactive power reserve and generators at buses 8 and 3 became critical. The generator at bus 8 was found more critical and nearer to Q-limits. At $k=$1.4, the generator 3 was the only one not exhausted and it was predicted to be critical. Fig. \ref{fig:14wvsi} shows the plot of WVSI and VSI for the most critical bus in the system.
%%\vspace{-0.0cm}
\begin{table}[h]
  \centering
  \caption{  IEEE 14 BUS SYSTEM CASE STUDY}
\begin{tabular}{ccccc}
\toprule
Loading Multiplier&$Q_T$&Predicted& $Q_{cr}$ &Actual\\
($k \times$ Base Case)&&Critical& &Critical\\
 &&Generators& &Generators\\
\cmidrule{1-5}
1.04 &$.7487$&  2& $.7794$& 2	\\
&& 8& $.85393$ &8\\
\cmidrule{1-5}
1.3 &$.8379$&  8 &$.84165$& 8	\\
&&3& $0.8926$ &3 \\
\cmidrule{1-5}
1.4 &$.8722$&  & & \\
&& 3& $.87932$& 3\\
\bottomrule
\end{tabular}
    \label{table:z_psse}
\end{table}
\subsubsection{IEEE 57 Bus System}
%%\vspace{-0.1cm}
\begin{table}
  \centering
  \caption{  IEEE 57 BUS SYSTEM CASE STUDY}
\begin{tabular}{ccccc}
\toprule
Loading Multiplier&$Q_T$&Predicted& $Q_{cr}$ &Actual\\
($k \times$ Base Case)&&Critical& &Critical\\
 &&Generators& &Generators\\
\cmidrule{1-5}
1.015 &$3.3845$&  9 &$3.4757$& 9	\\
&& 12& $3.682$ &12\\
\cmidrule{1-5}
1.3 &$3.7732$&  3 &$3.9733$& 3	\\
&&6 &$4.036$ &6 \\
\cmidrule{1-5}
1.5 &$4.046$&  & & \\
&& 8& $4.4777$& 8\\
\cmidrule{1-5}
1.6 &$4.1824$&  8& $4.4151$& 8	\\
&&2& $4.5406$ &2 \\
\cmidrule{1-5}
1.7 &$4.3324$&  8& $4.38$& 8	\\
&&2& $4.44$ &2 \\
\cmidrule{1-5}
1.72 &$4.3461$&  8& $4.377$& 8	\\
&&2& $4.4323$ &2 \\
\bottomrule
\end{tabular}
    \label{table:z_psse}
\end{table}
 At $k=$1.015, generators at buses 9 and 12 were predicted as critical generators. The generator at bus 9 was predicted to hit the Q-limits first closely followed by the generator at bus 12. This information matched the observed actual reactive power reserves from repeated power flows.\par
At the loading of $k=$1.3, the generators at buses 3 and 6 were predicted as critical generators. The generators at buses 9 and 12 were already at the Q-limits. After $k=$1.72 the power flow failed to converge. At this loading generators at buses 8 and 2 were the critical generators. Generator at bus 8 was predicted to hit the Q-limits first closely followed by the generator at bus 2. Fig. \ref{fig:57wvsi} shows the impact of Q-limits on bus 32 of IEEE 57 Bus system.
%\begin{figure}[htb]
 %\begin{center}
 %%\vspace{-0.4cm}
%\includegraphics[width=\columnwidth, height=6.0cm ]{IEEE57vsivswvsi2.eps}
%\caption{  \label{fig:57wvsi}Anticipating Impact of Q-limits [IEEE 57 Bus %System]}
 %\end{center}
%\end{figure}
\subsubsection{IEEE 118 Bus System}
 Proposed approach was correctly able to identify the generators to hit the Q-limits and thus include their impact when anticipating VSI. Preparing a list instead of predicting just one critical generator makes the process more robust when several generators could hit the Q-limits at almost same system load like $k=$2.0 for the IEEE 118 bus system when both generators at buses 104 and 105 were predicted to hit the Q-limits. Looking at actual power flow results the generator at bus 105 hits the Q-limit just before the generator at bus 104. Fig. \ref{fig:wvsi118} shows the impact of generator Q-limits on the most critical bus in IEEE-118 Bus system for this load increase scenario.
%%\vspace{-0.1cm}
\begin{table}[h]
  \centering
  \caption{  IEEE 118 BUS SYSTEM CASE STUDY}
\begin{tabular}{ccccc}
\toprule
Loading Multiplier&$Q_T$&Predicted& $Q_{cr}$ &Actual\\
($k \times$ Base Case)&&Critical& &Critical\\
 &&Generators& &Generators\\
\cmidrule{1-5}
1.01 &$14.432$&  12 &$16.034$& 12	\\
&& 70& $16.174$ &70\\
\cmidrule{1-5}
1.4 &$16.48$&77 &$16.508$ &77 \\
\cmidrule{1-5}
1.8 &$18.58$& 80& $18.965$& 80\\
\cmidrule{1-5}
2.0 &$19.63$ &  104& $20.231$& 104	\\
&&105& $20.231$ &105 \\
\cmidrule{1-5}
2.2 &$20.68$&  73& $20.98$& 73	\\
&&55& $21.187$ &55 \\
\cmidrule{1-5}
2.4 &$22.15$&  32& $22.689$& 32	\\
&&72& $22.832$ &72 \\
\bottomrule
\end{tabular}
    \label{table:z_psse}
\end{table}

%\begin{figure}[h]
 %\begin{flushleft}
%\includegraphics[width=\columnwidth, height=6.0cm ]{118vsivwvsi2.eps}
%\caption{  \label{fig:wvsi118}Anticipating Impact of Q-limits [IEEE 118 Bus %System]}
 %\end{flushleft}
%\end{figure}
%%\vspace{-0.5cm}
\subsection{Security Analysis}
%%%\vspace{-0.1cm}
The proposed approach for security analysis is demonstrated here for IEEE 14, IEEE 57 Bus and IEEE 118 Bus systems. The first step for security analysis is the prediction of contingency states using piecewise linear method. Once the post contingency states are known, perturbation analysis is performed to compute the voltage stability index. Criticality of contingency is decided based on the computed index.
%%\vspace{-.1cm}
\subsubsection{Classification of Contingencies}
In this subsection, the ability of the proposed scheme to classify contingencies accurately as critical or non-critical is demonstrated. Contingency classification capability of the proposed scheme is validated by forming a confusion matrix where actual labels come from continuation power flow and predicted labels come from the proposed approach. For WVSI based approach a threshold of 0.75 is used to classify critical and non-critical contingencies for IEEE 57 and IEEE 14 Bus systems.  $\Delta \lambda$ is the margin from the nose point in terms of per-unit apparent power. Contingencies having $\Delta \lambda >0.75$ are considered non-critical and contingencies having $\Delta \lambda <0.75$ are considered critical for IEEE 14 and 57 Bus System. For IEEE 118 Bus System a WVSI threshold of 0.85 and a $\Delta \lambda$ threshold of 0.9 is chosen to compute the confusion matrix. In this work threshold is chosen based on simulation experiments. Offline studies can be used to compute the threshold for different network configurations. Table 5 presents the confusion matrix to demonstrate the classification performance of proposed approach. Table 5 shows that out of 69 critical contingencies proposed approach is correctly able to identify 65 of them as critical. Similarly 446 out of 459 non-critical contingencies are identified correctly as non-critical. Table 6 presents the classification performance metrics using the proposed approach. Table 8 and Table 9 demonstrate the examples of security analysis for IEEE 14 and IEEE 57 Bus respectively.\\
%%\vspace{-0.3cm}
\begin{table}[h]
\centering
\caption{Confusion Matrix for Contingency Classification}
\begin{tabular}{llccc}
\toprule
%\multicolumn{2}{c}{Status}&\multicolumn{2}{c}{True}&\\
\cmidrule{1-5}
%\multicolumn{2}{c|}{}&Critical&Non %Critical&\multicolumn{1}{c}{}\\
&  & Critical & Non-Critical & \\
\cmidrule{1-5}
& \multicolumn{1}{c}{Critical (Predicted)} & $65$ & $13$ & \\
\cmidrule{1-5}
& \multicolumn{1}{c}{Non-Critical (Predicted)} & $04$ & $446$ & \\
\cmidrule{1-5}
\multicolumn{1}{c}{} & \multicolumn{1}{c}{Total} & \multicolumn{1}{c}{$69$} & \multicolumn{    1}{c}{$459$} & \multicolumn{1}{c}{}\\
\end{tabular}
\end{table}
%\vspace{-0.3cm}
\begin{table}[h]
  \centering
  \caption{Contingency Classification performance}
\begin{tabular}{cc}
\toprule
Metric&Value\\
\cmidrule{1-2}
Accuracy &  96.78\%\\
\cmidrule{1-2}
Precision &  83.33\% \\
\cmidrule{1-2}
Recall &  94.20\% \\
\cmidrule{1-2} 
F-score &  88.44\%\\
\cmidrule{1-2} 
\end{tabular}
\end{table}

%%\vspace{0.1cm}
\subsubsection{Ranking of Contingencies}
 In order to validate the ability of the proposed scheme to rank top 10 contingencies from a given list Wilcoxon signed rank test has been used. The actual ranks of contingencies come from the continuation power flow and predicted ranks come from the proposed scheme. Simulations were performed for IEEE 14 Bus, IEEE 57 Bus system and IEEE 118 Bus systems. In total 528 contingencies are simulated. 150 for IEEE 57 Bus system, 36 for IEEE 14 Bus system and 342 contingencies for IEEE 118 Bus system. 
 
 Results in Table 7 demonstrate that the proposed approach is adequately able to identify top 10 critical contingencies. Wilcoxon signed rank test confirms that same ranking trends are observed from the proposed index and the CPF based indices. A high p-value suggests that the null hypothesis (Median of difference between CPF and WVSI based rankings is zero) cannot be rejected thus validating the ranking capability of the proposed scheme.

\begin{table}[h]
  \centering
  \caption{  Ranking Performance: Wilcoxon Signed Rank Test}
\begin{tabular}{cccccc}
\toprule
Case&N& Loading &Confidence&p-value&Confidence\\
& & Multiplier& Interval& & \\
\cmidrule{1-6}
IEEE 14 &  10& 1& [-3.5, 8]&0.203& 95\%	\\
\cmidrule{1-6}
IEEE 14 &  10& 1.3& [-2, 3]&0.9453& 95\%	\\
\cmidrule{1-6}
IEEE 57 &  10& 1& [-5, 7.5]&0.5703& 95\%	\\
\cmidrule{1-6}
IEEE 57 &  10& 1.3& [-1, 5.5]&0.1152& 95\%	\\
\cmidrule{1-6}
IEEE 118 &  10& 1& [-4, 6.5]&0.3594& 95\%	\\
\cmidrule{1-6}
IEEE 118 &  10& 1.4& [-3, 4]&0.8457& 95\%	\\
\bottomrule
\end{tabular}
    \label{table:z_psse}
\end{table}

\begin{table}[h]
  \centering
  \caption{Example IEEE 14 Bus System Security Analysis}
\begin{tabular}{cccccc}
\toprule
\multirow{1}{*}{Contingency}&\multicolumn{2}{c}{Loading=1.0}&&\multicolumn{2}{c}{Loading=1.3}\\
\cmidrule{2-3}
\cmidrule{5-6}
&$WVSI_{pred}^{post}$ &Status&&$WVSI_{pred}^{post}$&Status \\ \hline
Br 5-6 & .381& NC&&.0.9286 &C \\ \hline %DelLamb=8e-4
Br 6-11 & .2116& NC&&.0.4517 &NC \\ \hline
Br 4-9 & .2322& NC&&.5472 &NC \\ \hline
Br 2-4 & .2040& NC&&.6499 &NC \\ 
\hline
Br 2-3 & .2534& NC&&.7626 &C \\  %DelLamb=.0024
\bottomrule
\end{tabular}
\label{table:sec14}
\end{table}

\begin{table}[h]
  \centering
  \caption{Example IEEE 57 Bus System Security Analysis}
\begin{tabular}{cccccc}
\toprule
\multirow{1}{*}{Contingency}&\multicolumn{2}{c}{Loading=1.0}&&\multicolumn{2}{c}{Loading=1.3}\\
\cmidrule{2-3}
\cmidrule{5-6}
&$WVSI_{pred}^{post}$ &Status&&$WVSI_{pred}^{post}$&Status \\ \hline
Br 13-15 & .5929& NC&&.8024 &C \\ \hline %DelLamb=0.595
Br 4-18 & .5784& NC&&.7388 &NC \\ \hline %DelLamb=.752
Br 52-53 & .5237& NC&&.7236 &NC \\ \hline %DelLamb=.548
Br 2-3 & .5499& NC&&.8182 &C \\ \hline %DelLamb=.0804
Br 3-4 & .4934& NC&&1.0586 &C \\ \hline %DelLamb=.0311
Br 4-5 & .5629& NC&&.7575 &C \\ \hline %DelLamb=.7413
Br 4-6 & .56& NC&&1.0092 &C \\ \hline %DelLamb=.6933
Br 5-6 & .5886& NC&&.7556 &C \\  %DelLamb=.8024
\bottomrule
\end{tabular}
\label{table:sec57}
\end{table}

%%\vspace{-0.6cm}
\section{Conclusions}
In this paper, an online hybrid voltage stability index is developed using synchrophasor data and with consideration of generator reactive power limits. Online security analysis is also performed using the developed voltage stability index. Critical generators are identified using PMU measurements and reactive power reserve information from the generator buses. With the knowledge of Voltage stability index at the existing operating point and voltage stability index for expected exhaustion of reactive power sources, operator can be better prepared to identify and deal with impending voltage stability problems. Once the critical generators are known, a modified Jacobian is constructed and used to compute a weighted Voltage stability index (WVSI). WVSI considers the present voltage stability status and reactive power reserve status in a single index. Security analysis to identify critical contingencies from a given list is performed by using the developed index. The proposed approach can provide timely alerts to the operator about the voltage stability problems in the system. Knowledge of critical generators can be utilized to plan effective control actions. As the deployment of PMUs in power system increase, linear state estimation is likely to become a key application for efficient power system monitoring. The proposed approach can be deployed at control centers to perform wide area monitoring using PMU based linear state estimators. The proposed approach can also work with SCADA based state estimation but at the slow reporting rate of voltage stability index. Developed approach has been validated for multiple test systems and shows superior performance compared to measurement based approach (e.g. estimation based Thevenin's approach). Future work will include benchmarking with CPF and implementation with real utility data.  
\section{Acknowledgements}
Authors would like to acknowledge the support from Fulbright scholarship to conduct this work. We would also like to thank Dr. Saugata Biswas for preliminary work.  
%%\vspace{-0.45cm}
%\bibliographystyle{iet}
%\bibliography{sample}
\bibliographystyle{iet}
\bibliography{sample}
\end{document}